\documentclass[sigconf,nonacm]{acmart}

\AtBeginDocument{%
  }

\usepackage{array}
\newcolumntype{P}[1]{>{\raggedright\arraybackslash}p{#1}}
\usepackage{tikz}
\usetikzlibrary{arrows.meta,positioning}

\newcommand{\tbl}[1]{\texttt{#1}}

\newcommand{\wt}[1]{\texttt{#1}}

\begin{document}

\title[WPTokens]{WPTokens: A Dataset for Studying the Regulation of Wikipedia}
\subtitle{Sentence- and node-level change tracking over the full revision
  history of French Wikipedia}

\author{L{\'e}o Joubert}
\email{leo.joubert@univ-rouen.fr}
\affiliation{%
  \institution{Universit{\'e} de Rouen Normandie, DySoLab}
  \city{Mont-Saint-Aignan}
  \country{France}
}

\renewcommand{\shortauthors}{L. Joubert}

\begin{abstract}
  Studying how Wikipedia is regulated requires knowing who adds, rewrites,
  moves and deletes which piece of content, when and where. WPTokens is a
  dataset that answers these questions over the full revision history of the
  French-language Wikipedia. It relies on a positional diff over two layers:
  a text layer made of sentences, and a node
  layer made of the structural elements of wikitext (templates, links,
  references, headings, tables, lists, etc.). Each piece of content keeps a
  stable identifier for its whole life, and every change to it is recorded as
  an addition, an update, a move, a move with update, or a deletion. Built
  from the April 2026 dump, the dataset covers 2.75 million articles, 153.6
  million revisions, 342.6 million tokens and 573.7 million token events,
  stored in an indexed SQLite database. A companion R package turns these
  records into temporal contributor--token multigraphs, completed by a layer
  of revert ties between contributors. We describe the tokenization, the diff
  algorithm, the data model and the graph construction, and discuss the
  limitations of the approach.
\end{abstract}




\maketitle

%% ===========================================================================
\section{Introduction}
%% ===========================================================================

Wikipedia has become one of the reference encyclopedias of Internet users.
Unlike a classical encyclopedia, where articles are written under the
supervision of an editorial board surrounded by experts, Wikipedia relies on
the mobilisation of a community of contributors who write through a
\emph{wiki}: a collaborative writing system in which every contribution is
layered upon the previous ones to form texts. Regulation is therefore not
located in a single place or moment; it is distributed across the stream of
edits through which contributors add content, rewrite it, move it around and
delete it. Deletions, in
particular, are traces of heterogeneous activities, from documentation and
sourcing to activism and surveillance.

A sociological analysis of this regulation first requires unfolding the
co-writing process, by asking questions that describe its flow: \emph{who adds
what, when and where?} and \emph{who deletes what, when and where?} WPTokens
answers them for the full history of French Wikipedia articles, with three
design choices. Content is followed as it is reworded and moved, so that
correcting a typo or relocating a paragraph is not mistaken for a deletion
followed by an addition. The structural layer of articles---references,
templates, categories, links, section headings---is kept, since much of the
regulation work takes place there: sourcing claims, tagging articles for
maintenance, categorising them. Finally, the unit of content is the sentence
or the whole structural element rather than the word, which keeps content
readable and the number of records manageable. Our contributions are:
\begin{itemize}
  \item a decomposition of wikitext into two layers, a text layer of sentences
    and a node layer of whole structural elements, computed by a single linear
    scan (Section~\ref{sec:tokenization});
  \item a positional diff that gives each piece of content a stable identifier
    and records its life as a sequence of typed events---addition, update,
    move, move with update, deletion (Section~\ref{sec:diff});
  \item the resulting dataset for the full history of French Wikipedia
    articles (Sections~\ref{sec:model} and~\ref{sec:stats});
  \item an R package that turns these records into temporal multigraphs
    linking contributors to tokens, completed by a layer of revert ties
    between contributors (Section~\ref{sec:graphs}).
\end{itemize}

%% ===========================================================================
\section{Scope and Source Data}
\label{sec:scope}
%% ===========================================================================

\paragraph{Source}
The dataset is built from the \emph{pages-meta-history} dump of French
Wikipedia of 1 April 2026, distributed as 165
compressed XML files containing the full wikitext of every revision.

\paragraph{Pages and revisions}
The focus is on the writing of encyclopedic content: only articles
(namespace~0) are processed, and redirect pages are skipped. Revisions whose
text or author has been suppressed are discarded. Talk pages are not
included.

\paragraph{Reverts}
A revision is \emph{reverting} if its content (SHA-1 hash) is identical to
that of one of the 15 preceding revisions, and the revisions in between are
\emph{reverted}. Both flags are stored with the revision metadata. Reverting
and reverted revisions are excluded from the content diff: a revert erases
other contributors' work in a single click, which does not, a priori, stem from the same social process as an addition or
a deletion made after reading. Diffs are thus computed on the \emph{clean}
sequence of the remaining revisions, so that an act of vandalism and its
cancellation leave no trace in the token tables. Reverts are not lost,
however: they remain in the revision table and form a dedicated layer of the
graphs (Section~\ref{sec:graphs}), so that this hypothesis can be put to the
test. Consecutive clean revisions with identical hashes carry no change and
are skipped.

%% ===========================================================================
\section{Tokenization in Two Layers}
\label{sec:tokenization}
%% ===========================================================================

Each revision is decomposed into two sequences of typed elements---the
\emph{tokens}---which are diffed independently: a \emph{node layer}, made of
the structural elements of the wikitext kept whole, and a \emph{text layer},
made of the sentences of the running prose. Both are produced by a single
left-to-right pass over the wikitext, without building a syntax tree. The two
layers overlap on purpose: the anchor text of a wikilink appears both in the
wikilink node and in the sentence that contains it, so that the survival of a
content can be followed whatever its markup.

\paragraph{Node layer}
HTML comments are first removed. Templates are then extracted by a
brace-matching scanner that keeps every outermost \wt{\{\{\ldots\}\}} span
whole, nested templates and parameters included, and ignores braces inside
comments and \wt{nowiki} blocks. Because this step comes first, a template
nested in another element---typically a citation template inside a
reference---is recorded as a template node in its own right and removed from
the enclosing element. The remaining text is scanned with a single regular
expression whose alternatives are tried in a fixed order of priority:
\wt{nowiki} blocks, HTML and extension tags (\wt{<ref>}, \wt{<gallery>},
\wt{<math>}, etc.), tables, headings, list items, media links, prefixed links
(interlanguage links, categories, portals), inline formatting tags,
wikilinks, external links and bare URLs. Since the first matching alternative
consumes its whole span, an element absorbs everything it contains: a
wikilink inside a reference belongs to the reference node and does not appear
on its own. Each match becomes a node whose text is the source span with
whitespace normalised; consecutive list items are merged into a single list
node. Table~\ref{tab:types} lists the resulting node types. In the node
sequence of a revision, templates come first, in order of appearance,
followed by the other nodes in order of appearance.

\begin{table}
  \caption{Token types. Only the elements marked \checkmark{} leave text in
    the prose from which sentences are built.}
  \label{tab:types}
  \small
  \begin{tabular}{@{}llc@{}}
    \toprule
    Type & Wikitext & Prose \\
    \midrule
    \multicolumn{3}{@{}l}{\emph{Text layer}} \\
    \tbl{sentence} & running text & --- \\
    \midrule
    \multicolumn{3}{@{}l}{\emph{Node layer}} \\
    \tbl{template}     & \wt{\{\{name|params\}\}} & \\
    \tbl{wikilink}     & \wt{[[target]]}, \wt{[[target|label]]} & \checkmark \\
    \tbl{media}        & \wt{[[File:\ldots]]}, \wt{[[Fichier:\ldots]]} & \\
    \tbl{interwiki}    & \wt{[[en:\ldots]]}, \wt{[[Cat{\'e}gorie:\ldots]]} & \\
    \tbl{externallink} & \wt{[http://\ldots{} label]}, bare URL & \checkmark \\
    \tbl{heading}      & \wt{== Title ==} & \\
    \tbl{tag}          & \wt{<ref>\ldots</ref>}, \wt{<i>\ldots</i>}, etc. & (\checkmark) \\
    \tbl{table}        & \wt{\{|\ldots|\}} & \\
    \tbl{list}         & consecutive \wt{*}, \wt{\#}, \wt{;}, \wt{:} lines & \\
    \bottomrule
  \end{tabular}

  \smallskip
  \raggedright\footnotesize (\checkmark) only inline formatting tags
  (\wt{<i>}, \wt{<b>}, \wt{<em>}, \wt{<strong>}, \wt{<tt>}).
\end{table}

\paragraph{Text layer}
The prose is what remains between matched elements, completed by the visible
text of some of them: the label (or, failing that, the target) of wikilinks,
the label of external links, and the content of inline formatting tags and
\wt{nowiki} blocks. Templates, other tags, tables, headings, lists, media and
prefixed links leave no text in the prose. Bold and italic quote markers are
removed. The prose is then segmented into sentences. It is first split into
paragraphs at blank lines; each paragraph is then split after a terminal
punctuation mark (\wt{.}, \wt{!}, \wt{?}) followed by whitespace and an
uppercase letter or a digit. Splits made after a common abbreviation (\wt{M},
\wt{Mme}, \wt{Dr}, \wt{vol}, \wt{p}, \wt{cf}, abbreviated months, etc.) are
undone. Stop words and punctuation are kept, so that sentences remain
readable and can be compared character by character.

\paragraph{Example}
Figure~\ref{fig:tokenization} shows the decomposition of a passage of the
article \emph{Nationalisme}. It yields
one sentence and eight nodes: three templates (one of which was nested in the
reference), four wikilinks and the reference itself. The century template
\wt{\{\{s-|XVIII\}\}} leaves a gap in the sentence (``\emph{à la fin du ,}''):
the content of templates is only tracked in the node layer
(Section~\ref{sec:limitations}).

\begin{figure*}
  \small
  \begin{minipage}[t]{\textwidth}
    \textbf{(a) Wikitext.}\par\smallskip
    \raggedright\ttfamily\footnotesize
    Le '{}'{}'nationalisme'{}'{}' est un principe politique apparu à la fin du
    \{\{s-|XVIII\}\}, tendant à légitimer l'existence d'un [[État-nation]] pour
    chaque [[peuple]], défini par des caractéristiques propres et communes à ses
    membres, comme une [[langue]], des traditions historiques et culturelles ou
    des [[Valeur (sociologie)|valeurs politiques]]<ref>\{\{harvsp|Gellner|1989
    (1983 pour l'édition anglaise)|loc=chapitre 1\}\}. Le terme de
    '{}'principe politique'{}' est repris dans
    \{\{harvsp|Hobsbawm|1992|loc=chapitre I\}\} et revendiqué comme venant du
    texte de Ernest Gellner.</ref>.
  \end{minipage}

  \bigskip
  \begin{minipage}[t]{0.49\textwidth}
    \textbf{(b) Text layer (1 sentence).}\par\smallskip
    \raggedright\footnotesize
    \begin{tabular}{@{}lP{0.85\textwidth}@{}}
      \tbl{sentence} & \ttfamily Le nationalisme est un principe politique
      apparu à la fin du , tendant à légitimer l'existence d'un État-nation
      pour chaque peuple, défini par des caractéristiques propres et communes
      à ses membres, comme une langue, des traditions historiques et
      culturelles ou des valeurs politiques. \\
    \end{tabular}
  \end{minipage}\hfill
  \begin{minipage}[t]{0.49\textwidth}
    \textbf{(c) Node layer (8 nodes).}\par\smallskip
    \raggedright\footnotesize
    \begin{tabular}{@{}lP{0.75\textwidth}@{}}
      \tbl{template} & \ttfamily \{\{s-|XVIII\}\} \\
      \tbl{template} & \ttfamily \{\{harvsp|Gellner|1989 (1983 pour l'édition
                       anglaise)|loc=chapitre 1\}\} \\
      \tbl{template} & \ttfamily \{\{harvsp|Hobsbawm|1992|loc=chapitre I\}\} \\
      \tbl{wikilink} & \ttfamily [[État-nation]] \\
      \tbl{wikilink} & \ttfamily [[peuple]] \\
      \tbl{wikilink} & \ttfamily [[langue]] \\
      \tbl{wikilink} & \ttfamily [[Valeur (sociologie)|valeurs politiques]] \\
      \tbl{tag}      & \ttfamily <ref>. Le terme de '{}'principe
                       politique'{}' est repris dans et revendiqué comme venant
                       du texte de Ernest Gellner.</ref> \\
    \end{tabular}
  \end{minipage}
  \caption{Decomposition of a passage of the article \emph{Nationalisme}. (a)
    Source wikitext. (b) Text layer. (c) Node layer, in the order of the
    sequence (templates first).}
  \Description{A paragraph of wikitext from the French article on
    nationalism, followed by its decomposition into one sentence and eight
    structural nodes (three templates, four wikilinks and one reference tag).}
  \label{fig:tokenization}
\end{figure*}

%% ===========================================================================
\section{Positional Diff}
\label{sec:diff}
%% ===========================================================================

For a page, let $r_1, \ldots, r_n$ be its clean revisions in chronological
order and, for a given layer, let $S_k$ be the sequence of typed elements of
$r_k$. The diff from $r_{k-1}$ to $r_k$ is computed independently on each
layer. Each element of $S_k$ either inherits the identifier of an element of
$S_{k-1}$ or receives a new one; an identifier is created only when a content
appears and closed only when it disappears.

\paragraph{Alignment}
$S_{k-1}$ and $S_k$ are first aligned as sequences of (text, type) pairs,
after trimming their common prefix and suffix, with the Ratcliff--Obershelp
algorithm. The alignment yields four kinds of blocks: \emph{equal}
blocks, whose elements are unchanged and in place and inherit their
identifier; \emph{replace} blocks, which pair a run of old elements with a run
of new ones at the same place; and \emph{delete} and \emph{insert} blocks,
which contain old and new elements without counterpart.

\paragraph{Similarity and matching}
Two elements $a$ and $b$ are considered the same content if they have the
same type and $d(a,b) \leq \tau$, where
\[
  d(a,b) = 1 - \frac{2\,M(a,b)}{|a| + |b|}
\]
is the character-level Ratcliff--Obershelp distance, $M(a,b)$ being the number
of matched characters. The dataset uses $\tau = 0.5$ for both layers. Within a set of candidate old and new elements,
pairs are formed greedily: all pairs at distance at most $\tau$ are sorted by
increasing distance and accepted when neither element is already paired.

\paragraph{Three phases}
\emph{Phase~1 (in place).} Matching is run within each replace block. Paired
elements keep the old identifier; if $d > 0$, an \emph{update} event is
recorded.
\emph{Phase~2 (relocated).} The old elements still unpaired---from delete
blocks and from phase~1---are matched with the new elements still
unpaired---from insert blocks and from phase~1---across the whole revision.
Paired elements keep the old identifier. If their position changed, a
\emph{move} event is recorded, or a \emph{move with update} if $d > 0$;
otherwise, an update is recorded if $d > 0$.
\emph{Phase~3 (deaths and births).} The remaining old elements are
\emph{deleted}: their lifecycle is closed at $r_k$. The remaining new
elements are \emph{added} and receive a fresh identifier.

\paragraph{Positions}
Each event records the position of the element before and after the change:
its index in the layer, the number of elements in the layer, and the
normalised index $i/(N-1)$. Since only relocated elements produce moves, the
passive drift of positions---an element shifting because others were inserted
or deleted before it---is never recorded as a move. For an update, the before
and after positions are both set to the current index.

Table~\ref{tab:events} summarises the five event types and how they are
encoded. Table~\ref{tab:toy} runs the algorithm on a toy page edited by three
contributors.

\begin{table}
  \caption{Event types and their encoding.}
  \label{tab:events}
  \small
  \begin{tabular}{@{}llll@{}}
    \toprule
    Event & Table & \tbl{position\_from} & \tbl{text} \\
    \midrule
    addition    & \tbl{token\_events} & \tbl{NULL}            & initial text \\
    update      & \tbl{token\_events} & $=$ \tbl{position\_to}    & new text \\
    move        & \tbl{token\_events} & $\neq$ \tbl{position\_to} & \tbl{NULL} \\
    move+update & \tbl{token\_events} & $\neq$ \tbl{position\_to} & new text \\
    deletion    & \tbl{tokens\_map}   & \multicolumn{2}{l}{\tbl{rev\_deleted} set} \\
    \bottomrule
  \end{tabular}
\end{table}

\begin{table*}
  \caption{The diff on a toy page. In $r_2$, Bob rewrites the first sentence
    and removes the link on \emph{langue}; in $r_3$, Carol puts the second
    sentence first and adds a new one. Identifiers are shared by both layers
    of a page. Sentence $t_2$ is untouched by $r_2$: the removal of the link
    is visible in the node layer only.}
  \label{tab:toy}
  \small
  \begin{tabular}{@{}lllP{0.62\textwidth}@{}}
    \toprule
    Revision & Token & Event (positions) & Text \\
    \midrule
    $r_1$ (Alice) & \multicolumn{3}{@{}p{0.86\textwidth}@{}}{\raggedright\ttfamily\footnotesize
      Le '{}'{}'nationalisme'{}'{}' est un principe politique. Il légitime
      l'existence de l'[[État-nation]]. Il s'appuie sur une [[langue]] commune.} \\
     & $t_0$ sentence & addition ($\cdot \to 0$) & Le nationalisme est un principe politique. \\
     & $t_1$ sentence & addition ($\cdot \to 1$) & Il légitime l'existence de l'État-nation. \\
     & $t_2$ sentence & addition ($\cdot \to 2$) & Il s'appuie sur une langue commune. \\
     & $t_3$ wikilink & addition ($\cdot \to 0$) & \texttt{[[État-nation]]} \\
     & $t_4$ wikilink & addition ($\cdot \to 1$) & \texttt{[[langue]]} \\
    \midrule
    $r_2$ (Bob) & $t_0$ sentence & update ($0 \to 0$), $d = 0.14$ & Le nationalisme est une idéologie politique. \\
     & $t_4$ wikilink & deletion & --- \\
    \midrule
    $r_3$ (Carol) & $t_1$ sentence & move ($1 \to 0$) & \tbl{NULL} \\
     & $t_5$ sentence & addition ($\cdot \to 3$) & Il apparaît à la fin du XVIIIe siècle. \\
    \bottomrule
  \end{tabular}
\end{table*}

%% ===========================================================================
\section{Data Model}
\label{sec:model}
%% ===========================================================================

The dataset is a single SQLite database, whose five tables are described in
Table~\ref{tab:schema}.

\begin{table*}
  \caption{Tables of the database. Keys are in italics.}
  \label{tab:schema}
  \small
  \begin{tabular}{@{}lP{0.42\textwidth}P{0.38\textwidth}@{}}
    \toprule
    Table & Columns & One row per\ldots \\
    \midrule
    \tbl{revisions} & \emph{rev\_id}, page\_id, page\_title,
      page\_namespace, parent\_id, timestamp, user, user\_id, sha1, bytes,
      minor, revert\_is\_reverting, revert\_reverted\_by,
      revert\_reverted\_to\_by
      & revision, with its metadata and revert flags (all revisions, including
      reverting and reverted ones). \\
    \tbl{tokens\_map} & \emph{page\_id, token\_id}, token\_type, rev\_added,
      rev\_deleted
      & token: identity, type and lifecycle; \tbl{rev\_deleted} is
      \tbl{NULL} if the token is still present at the end of the dump. \\
    \tbl{token\_events} & page\_id, token\_id, rev\_id, text, position\_from,
      position\_from\_norm, position\_from\_total, position\_to,
      position\_to\_norm, position\_to\_total
      & addition, update, move or move with update of a token
      (Table~\ref{tab:events}). \\
    \tbl{sample\_pages} & page\_id, page\_namespace, draw\_order, n\_edits
      & page, in a random order fixed per namespace. \\
    \tbl{sample\_users} & user\_id, user, page\_namespace, draw\_order, n\_edits
      & contributor and namespace, in a random order fixed per namespace. \\
    \bottomrule
  \end{tabular}
\end{table*}

\paragraph{Identity and text}
A token is identified by the pair (\tbl{page\_id}, \tbl{token\_id}): the
counter restarts on each page and is shared by both layers.
\tbl{tokens\_map} is a pure identity card and holds no text. Text lives in
\tbl{token\_events} and is stored only when content appears or changes; it is
\tbl{NULL} for pure moves. The text of a token at any revision is thus
obtained by carrying forward its last non-null text. Deletions are not
events: they are read from \tbl{tokens\_map.rev\_deleted}.

\paragraph{Joins and aggregates}
Both token tables join \tbl{revisions} on revision identifiers
(\tbl{rev\_added}, \tbl{rev\_deleted}, \tbl{rev\_id}), which gives the author
and timestamp of every event. Anonymous contributors have a \tbl{NULL}
\tbl{user\_id} and are identified by their IP address in \tbl{user}. There is
no precomputed table of edits: the number of additions, updates, moves and
deletions of a revision is computed on demand by aggregating the two token
tables. Indices on pages, contributors, revisions and tokens make it possible
to retrieve the full history of a page or of a contributor without scanning
the tables.

\paragraph{Sampling}
\tbl{sample\_pages} and \tbl{sample\_users} list pages and contributors with
their number of revisions, in a random order drawn once per namespace.
Drawing a uniform sample of size $n$ amounts to selecting
$\tbl{draw\_order} \leq n$, possibly within strata of activity defined on
\tbl{n\_edits}.

%% ===========================================================================
\section{Contributor--Token Graphs}
\label{sec:graphs}
%% ===========================================================================

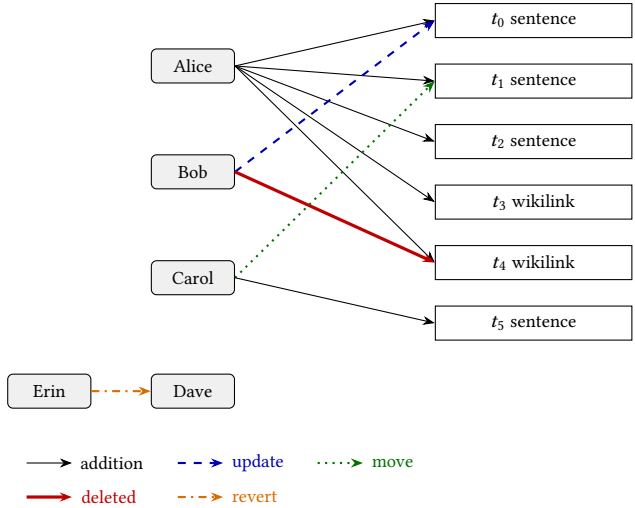
\begin{figure}
  \centering
  \begin{tikzpicture}[
      font=\footnotesize,
      contrib/.style={draw, rounded corners=2pt, fill=gray!12,
        minimum width=11mm, minimum height=4.5mm, inner sep=1pt},
      token/.style={draw, minimum width=26mm, minimum height=4.5mm,
        inner sep=1pt, anchor=west},
      add/.style={-{Stealth[length=1.6mm]}, black},
      upd/.style={-{Stealth[length=1.6mm]}, blue!70!black, dashed, thick},
      mov/.style={-{Stealth[length=1.6mm]}, green!45!black, dotted, thick},
      del/.style={-{Stealth[length=1.6mm]}, red!75!black, very thick},
      rev/.style={-{Stealth[length=1.6mm]}, orange!85!black, thick,
        dash dot},
    ]
    \node[contrib] (A) at (0, 4.0) {Alice};
    \node[contrib] (B) at (0, 2.6) {Bob};
    \node[contrib] (C) at (0, 1.2) {Carol};
    \node[contrib] (D) at (0, -0.3) {Dave};
    \node[contrib] (E) at (-1.9, -0.3) {Erin};

    \node[token] (t0) at (3.2, 4.6) {$t_0$ sentence};
    \node[token] (t1) at (3.2, 3.8) {$t_1$ sentence};
    \node[token] (t2) at (3.2, 3.0) {$t_2$ sentence};
    \node[token] (t3) at (3.2, 2.2) {$t_3$ wikilink};
    \node[token] (t4) at (3.2, 1.4) {$t_4$ wikilink};
    \node[token] (t5) at (3.2, 0.6) {$t_5$ sentence};

    \foreach \t in {t0, t1, t2, t3, t4}
      \draw[add] (A.east) -- (\t.west);
    \draw[upd] (B.east) -- (t0.west);
    \draw[del] (B.east) -- (t4.west);
    \draw[mov] (C.east) -- (t1.west);
    \draw[add] (C.east) -- (t5.west);
    \draw[rev] (E.east) -- (D.west);

    \begin{scope}[shift={(-2.2,-1.25)}]
      \draw[add] (0,0) -- (0.6,0) node[right] {addition};
      \draw[upd] (2.0,0) -- (2.6,0) node[right] {update};
      \draw[mov] (3.85,0) -- (4.45,0) node[right] {move};
      \draw[del] (0,-0.45) -- (0.6,-0.45) node[right] {deleted};
      \draw[rev] (2.0,-0.45) -- (2.6,-0.45) node[right] {revert};
    \end{scope}
  \end{tikzpicture}
  \caption{Contributor--token graph of the toy page of Table~\ref{tab:toy},
    to which a vandalism by Dave ($r_4$), reverted by Erin ($r_5$), has been
    added. These two revisions are excluded from the token layer and only
    appear as a revert edge.}
  \Description{A directed graph with contributors on the left and six tokens
    on the right. Alice has addition edges to tokens t0 to t4; Bob has an
    update edge to t0 and a deletion edge to t4; Carol has a move edge to t1
    and an addition edge to t5; Erin has a revert edge to Dave.}
  \label{fig:graph}
\end{figure}

\paragraph{Definition}
For a page $p$, let $C_p$ be the set of its contributors and $T_p$ the set
of its tokens. We define a directed, temporal multigraph
$G_p = (C_p \cup T_p,\ E^{\mathrm{tok}}_p \cup E^{\mathrm{rev}}_p)$ with two
layers of edges:
\begin{itemize}
  \item $E^{\mathrm{tok}}_p$ holds one edge $(c, t, \lambda, \theta)$ for each
    event on token $t$, from the author $c$ of the revision in which it
    occurred, labelled with its type $\lambda$ (addition, update, move,
    move\_update or deleted) and timestamped with the date $\theta$ of the
    revision;
  \item $E^{\mathrm{rev}}_p$ holds one edge $(c, c', \text{revert}, \theta)$
    each time a revision by $c$ reverts a revision by $c'$, self-reverts
    excluded.
\end{itemize}
There are no edges between tokens. Each event gives its own edge, so that
parallel edges coexist, and nodes are scoped to the page: a contributor
active on two pages yields two nodes, and the graphs of different pages are
disjoint. Figure~\ref{fig:graph} shows the graph of the toy page of
Table~\ref{tab:toy}, completed by an act of vandalism and its revert.

\paragraph{Uses}
The token layer records interactions mediated by content: two contributors
are related when they act on the same token, and the sequence of typed,
timestamped edges incident to a token tells its story---added by one
contributor, reworded by a second, deleted by a third. Classical interaction
networks are projections of $G_p$. Linking $u$ to $v$ whenever $u$ deleted a
token added by $v$ gives a deletion network; update networks are obtained in the
same way, and all projections can be restricted to some token types---only
sentences, or only references and templates. The revert layer, built from the
revisions that the token layer leaves out, captures the other mode of
erasure.

\paragraph{Implementation}
The R package \texttt{comsav} provides client-side access to the database.
Its function \texttt{comsav\_page\_network()} builds $G_p$ for one or several
pages, as R graph objects or as a SQLite file with four tables---\tbl{nodes},
\tbl{edges}, \tbl{tokens} and \tbl{page}. Its \tbl{tokens} table gives, for each token edge, the text of the
token before and after the event. The package also retrieves the revisions of
pages or contributors, reconstructs the history of a token, computes the
per-revision counts of events, draws samples, and implements metrics such as
the burstiness of contributors' activity or the controversy measure~$M$. On the article \emph{Miss France
2000}, included in the example database shipped with the package (98
revisions), the graph has 61 contributors, 461 tokens and 902 edges: 461
additions, 168 updates, 6 moves, 4 moves with update, 261 deletions and 2
reverts.

%% ===========================================================================
\section{Descriptive Statistics}
\label{sec:stats}
%% ===========================================================================

\paragraph{Revisions and contributors}
The dataset covers 2,748,733 articles and 153,561,219 revisions made between
4 August 2001 and 1 April 2026, that is 56 revisions per article on average
(Table~\ref{tab:stats}). Unregistered contributors, identified by their IP
address, made 15.4\% of the revisions. 6,403,319 revisions (4.2\%) are
reverting and 8,211,752 (5.3\%) are reverted; together, 13,822,141 revisions
(9.0\%) are left out of the content diff.

\begin{table}
  \caption{Main figures of the dataset (articles of French Wikipedia).}
  \label{tab:stats}
  \small
  \begin{tabular}{@{}lr@{}}
    \toprule
    Articles                      & 2,748,733 \\
    Revisions                     & 153,561,219 \\
    \quad reverting               & 6,403,319 \\
    \quad reverted                & 8,211,752 \\
    Contributors (registered)     & 1,284,728 \\
    Contributors (IP addresses)   & 7,268,712 \\
    Tokens                        & 342,580,655 \\
    Token events                  & 573,681,722 \\
    Database size                 & $\approx$304\,GB \\
    \bottomrule
  \end{tabular}
\end{table}

\paragraph{Tokens}
Templates are the most frequent type of token (35.3\%), ahead of wikilinks
(21.4\%) and sentences (18.9\%): four tokens out of five belong to the node
layer (Table~\ref{tab:tokentypes}). 65.6\%
of tokens are still present at the end of the dump. Survival varies by type:
headings are the most durable (79.9\%), and sentences (60.6\%) are deleted
more often than templates and wikilinks. Interlanguage links and categories
(50.4\%) probably bear the trace of the mass removal of interlanguage links
when they moved to Wikidata in 2013. External links outside references are
almost all deleted (9.0\% still present), which is consistent with editorial
guidelines that discourage such links in the body of articles.

\begin{table}
  \caption{Tokens by type, and share still present at the end of the dump.}
  \label{tab:tokentypes}
  \small
  \begin{tabular}{@{}lrrr@{}}
    \toprule
    Type & Tokens & Share (\%) & Present (\%) \\
    \midrule
    \tbl{template}     & 120,939,937 & 35.3 & 69.7 \\
    \tbl{wikilink}     &  73,437,453 & 21.4 & 68.4 \\
    \tbl{sentence}     &  64,742,439 & 18.9 & 60.6 \\
    \tbl{interwiki}    &  24,107,991 &  7.0 & 50.4 \\
    \tbl{heading}      &  22,373,030 &  6.5 & 79.9 \\
    \tbl{tag}          &  22,293,330 &  6.5 & 58.3 \\
    \tbl{list}         &   9,512,013 &  2.8 & 57.1 \\
    \tbl{media}        &   2,651,562 &  0.8 & 52.5 \\
    \tbl{table}        &   1,699,768 &  0.5 & 60.6 \\
    \tbl{externallink} &     823,132 &  0.2 &  9.0 \\
    \midrule
    Total              & 342,580,655 & 100.0 & 65.6 \\
    \bottomrule
  \end{tabular}
\end{table}

\paragraph{Events}
Each token has exactly one addition, and 117,884,553 tokens have been
deleted. The 573,681,722 rows of \tbl{token\_events} thus comprise
342,580,655 additions and 231,101,067 updates and moves---0.67 per token on
average.

%% ===========================================================================
\section{Availability and Reuse}
\label{sec:availability}
%% ===========================================================================

\paragraph{Data}
The database is developed and analysed by the COMSAV team, within the
DySoLab laboratory of the Universit{\'e} de Rouen Normandie. To join the team,
or to propose a research project based on the database, write to its
coordinator, L{\'e}o Joubert
(\href{mailto:leo.joubert@univ-rouen.fr}{leo.joubert@univ-rouen.fr}).

\paragraph{Graphs}
The R package \texttt{comsav} (Section~\ref{sec:graphs}) is available at
\url{https://gitlab.huma-num.fr/comsav/comsav}. It ships with a test
database of ten articles (940 revisions) in the same format, which makes it
possible to try the package without the full dataset.

%% ===========================================================================
\section{Limitations}
\label{sec:limitations}
%% ===========================================================================

\paragraph{Coverage}
Built from a dump, the dataset ignores pages deleted before the dump and
revisions whose content has been suppressed. It covers the articles of French
Wikipedia only.

\paragraph{Tokenization}
The scanner recognises wikitext elements without rendering them. Templates
that display text---dates, centuries, units, foreign-language
markers---leave gaps in sentences, and their content is only tracked in the
node layer. The content of tables and lists does not enter the text layer
either, which matters for list-heavy articles. Since templates are listed
before the other nodes, node positions do not reflect the interleaving of
templates and other elements. When a revision contains unbalanced braces, no
template is extracted from it and its templates fall into the prose, so that
the next well-formed revision appears as a large rewrite. Sentence
segmentation splits after initials (\emph{J.~S.~Bach}) and after unlisted
abbreviations, and misses boundaries followed by a lowercase letter or a
quotation mark.

\paragraph{Diff}
The thresholds $\tau = 0.5$ have not yet been calibrated against annotated
data; a sensitivity analysis is planned. Matching is greedy and one-to-one:
splitting a sentence in two yields an update and an addition, merging two
sentences an update and a deletion. Identifiers are local to a page and
never reused: content moved to another page, or deleted and later
reintroduced, receives a new identifier.

\paragraph{Reverts}
Only identity reverts within a window of 15 revisions are detected: partial
reverts and manual cancellations are treated as ordinary
edits. Conversely, excluding reverted revisions also
removes good-faith edits that were reverted wholesale, together with any
content they added.

\paragraph{Storage}
A single SQLite file of about 300\,GB gives fast access to a page or a
contributor through its indices, but full scans---counting events by type
over the whole dataset, for instance---take hours. Columnar formats could
ease the distribution and the analysis of the whole dataset.

%% ===========================================================================
\section{Conclusion}
%% ===========================================================================

WPTokens describes the full history of French
Wikipedia articles as the lives of 342.6 million pieces of content---sentences
and structural elements---, each followed from its addition to its possible
deletion through its updates and moves, and attributed to the contributors
who acted on it. Together with the reverts, these records can be turned into
temporal contributor--token graphs, from which classical interaction
networks are derived as projections. Future work includes the calibration of
the matching thresholds, the extension to talk pages and to other language
editions of Wikipedia, which would allow comparing their regulation, and the
distribution of the data in columnar formats.

\begin{acks}
This work was supported by the French National Research Agency (ANR) through
the COMSAV project (ANR-25-CE38-7350).
\end{acks}

\end{document}